\documentclass[preprints,article,accept,moreauthors,pdftex]{Definitions/mdpi}
\firstpage{1}
\pubvolume{1}
\issuenum{1}
\articlenumber{0}
\pubyear{XXXX}
\copyrightyear{XXXX}
\datereceived{XX Month Year}
\dateaccepted{XX Month Year}
\datepublished{}
\hreflink{https://doi.org/} 
\pdfoutput=1
\Title{Quantum-Inspired Neural Network Model of Optical Illusions}

\TitleCitation{Quantum-Inspired Neural Network Model of Optical Illusions}

\definecolor{my_green}{rgb}{0.55, 0.71, 0.0}

\newcommand{\icol}[1]{
  \left[\begin{smallmatrix}#1\end{smallmatrix}\right]%
}

\usepackage{textcomp}
\usepackage{bm}
\usepackage{comment}

\Author{Ivan S.~Maksymov \orcidA{}}

\AuthorNames{Ivan S.~Maksymov}

\AuthorCitation{Maksymov, I.S.}

\address{%
\quad Artificial Intelligence and Cyber Futures Institute, Charles Sturt University, Bathurst, NSW 2795, Australia; imaksymov@csu.edu.au}

\abstract{Ambiguous optical illusions have been a paradigmatic object of fascination, research and inspiration in arts, psychology and video games. However, accurate computational models of perception of ambiguous figures have been elusive. In this paper, we design and train a deep neural network model to simulate the human's perception of the Necker cube, an ambiguous drawing with several alternating possible interpretations. Defining the weights of the neural network connection using a quantum generator of truly random numbers, in agreement with the emerging concepts of quantum artificial intelligence and quantum cognition we reveal that the actual perceptual state of the Necker cube is a qubit-like superposition of the two fundamental perceptual states predicted by classical theories. Our results will find applications in video games and virtual reality systems employed for training of astronauts and operators of unmanned aerial vehicles. They will also be useful for researchers working in the fields of machine learning and vision, psychology of perception and quantum-mechanical models of human mind and decision-making.}


\begin{document}
\section{Introduction\label{sec:1}}
Optical illusions have fascinated humans since the ancient times \cite{Sha17, Nek32, Was34} and served as both object of inspiration in arts \cite{Fis67, Lin01} and paradigmatic topic of research in the fields of psychology and behavioural science \cite{Lon04, Kor05, Con09, Bus12, Sto12, Kor12, Run16, Mei19}. Nowadays, when artificial intelligence (AI) is all around, a question arises whether a computer or robotic system can recognise optical illusions similarly to a human. Apart from a blue-skies research goal of creating a humanoid robot that both aesthetically resembles a human and can perceive the world as a human, a practicable model of human perception of optical illusions could revolutionise the way video games and spatial computing systems are designed \cite{Mak23_quantum}, psychiatric illnesses are studied \cite{Bas16} and the effect of gravitation on cognition is investigated \cite{Yam06, Cle17}. Moreover, establishing the psychological and physiological origin of perception of ambiguous figures promises to unlock the secrets of human decision-making \cite{Bus12}, also revealing a long-hypothesised but yet elusive link between human mental states and quantum mechanics \cite{Khr06, Bus12, mindell2012quantum, wendt2015quantum, Con09, Atm10, Aer22, Kau22}.

\begin{figure}[h]
 \includegraphics[width=0.95\textwidth]{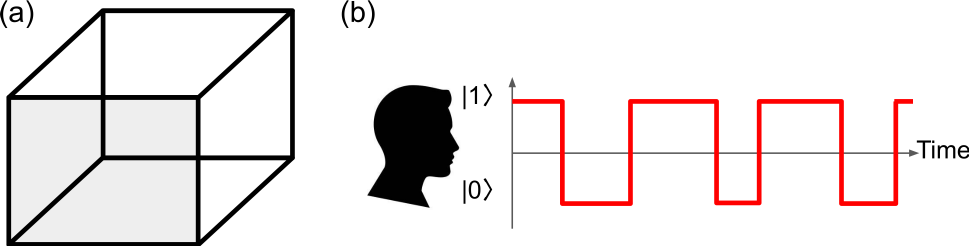}
 \caption{{\bf(a)}~The Necker cube. The answer to the question `Is the shaded face of the cube at the front or at the rear?' will change suddenly depending on the observer's perception, giving rise to a series of rectangular pulses corresponding to the front, $|0\rangle$, and rear, $|1\rangle$, perceptual states of the cube shown in panel~{\bf(b)}.\label{Fig1}}
\end{figure}

Consider the Necker cube \cite{Nek32} shown in Figure~\ref{Fig1}a. A simple self-examination aimed to answer the question `Is the shaded face of the cube at the front or at the rear?' will result in a series of possible interpretations that randomly switch between `front' and 'rear'. When the sequence of the observed front-rear states, which in the following we denote as $|0\rangle$ and $|1\rangle$, is recorded as a function of time as schematically shown in Figure~\ref{Fig1}b, one obtains a signal consisting of rectangular pulses of random duration. The temporal dynamics of the pulses will vary from one self-examination to another since the perception of optical illusions depends on the observer's age and gender \cite{lo2011investigation}. However, the general trend illustrated in Figure~\ref{Fig1}b will be similar for all observers. 

On the other hand, electroencephalograms recorded consistently with subjective inputs given by observers of the Necker cube and other ambiguous figures \cite{Gae98, Pia17, Joo20} suggest that the perception does not undergo an abrupt switching as sketched in Figure~\ref{Fig1}c but exhibits a rather continuous oscillation-like behaviour between the $|0\rangle$ and $|1\rangle$ states. Data speaking in favour of such a behaviour were also obtained in eye-tracking experiments, where both blink and movement of eye were associated with a perceptual reversal \cite{Lon04, Cho20, Mat23}. 

These experimental results indicate that the perceptual state may not exactly be $|0\rangle$ or $|1\rangle$ state but their combination. Mathematically, this scenario can be described as a superposition of $|0\rangle$ and $|1\rangle$ \cite{Bus12}. This intriguing observation has motivated the attempts to apply the methods of quantum mechanics and quantum computing to the analysis of human perception \cite{Atm10, Atm13, Bus12, Pot22}.    

A quantum computer uses a quantum bit (qubit) that can be in the states $|0 \rangle = \icol{1\\0}$ and $|1 \rangle = \icol{0\\1}$. These states are analogous to the `0' and `1' binary states of a classical digital computer. However, a qubit exists in a continuum of states between $|0 \rangle$ and $|1 \rangle$, i.e.~its states are a superposition $|\psi\rangle = \alpha |0 \rangle + \beta |1 \rangle$ with $|\alpha|^2 + |\beta|^2 = 1$.

\begin{figure}[h]
 \includegraphics[width=0.99\textwidth]{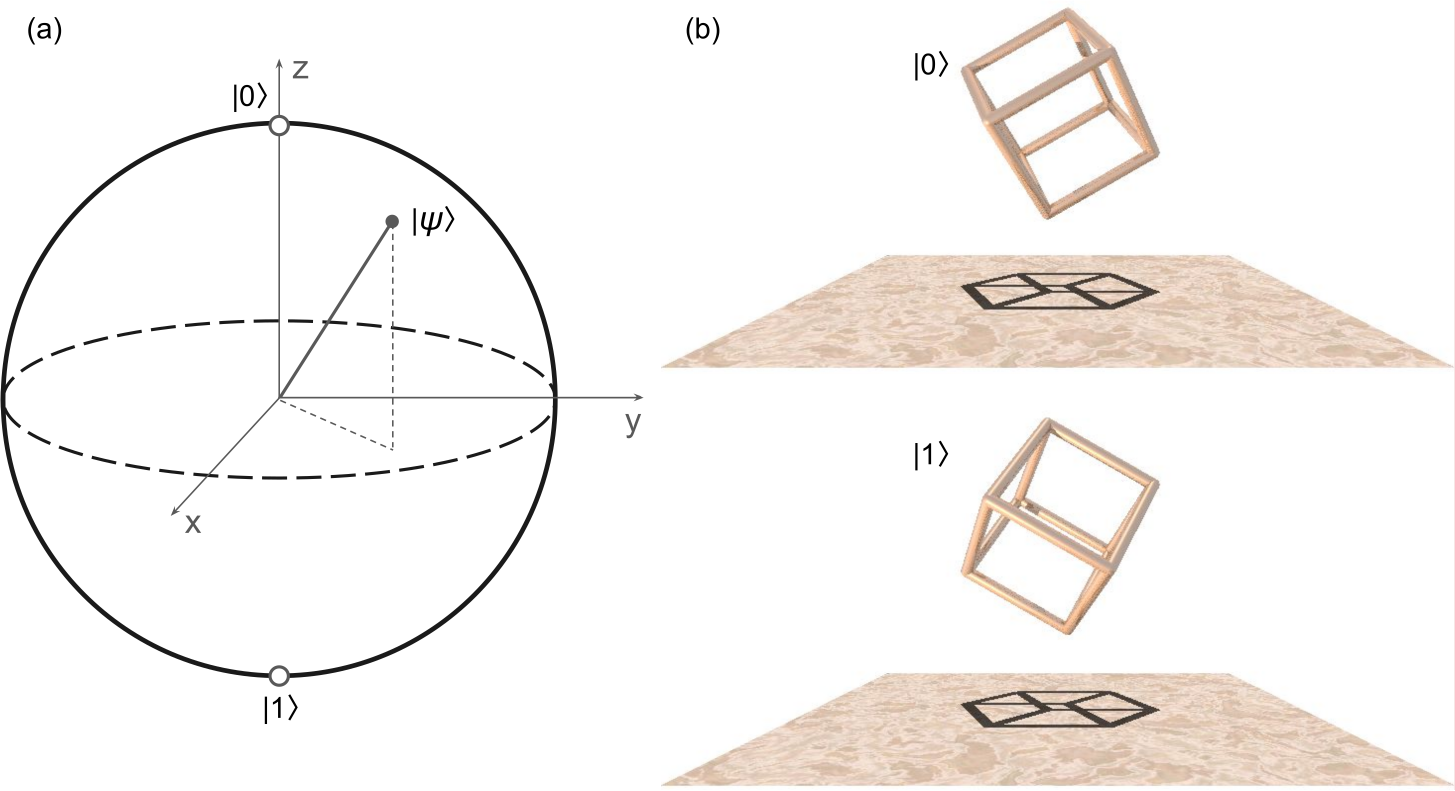}
 \caption{{\bf(a)}~Projective measurement of a qubit. {\bf(b)}~Projective qubit-like measurement applied to the Necker cube. The two-dimensional shadows of the cubes are identical and perceived by an observer as an ambiguous Necker cube. Considering the shadows as a qubit-like superposition of $|0\rangle$ and $|1\rangle$, we virtually project the shadows back to the three-dimensional space to obtain an unambiguous cube that corresponds to one of the basis states.\label{Fig1_1}}
\end{figure}

When a quantum measurement is done, a closed qubit system interacts in a controlled way with an external system from which the state of the qubit under measurement can be recovered. For example, using the projective measurement operators $M_0 = |0\rangle\langle0|$ and $M_1 = |1\rangle\langle1|$ \cite{Nie02}, the measurement probabilities for $|\psi\rangle = \alpha |0 \rangle + \beta |1 \rangle$ are $P_{|0\rangle} = |\alpha|^2$ and $P_{|1\rangle} = |\beta|^2$, which means that the qubit will be in one of its basis states. Such a projective measurement can be visualised using the concept of the Bloch sphere where the qubit is projected on one of the coordinate axes (e.g., $z$-axis in Figure~\ref{Fig1_1}a). 

Computational algorithms based on measurements of the states of a qubit are exponentially faster than any possible deterministic classical algorithm \cite{Nie02}. Subsequently, it has been demonstrated that quantum mechanics can explain certain psychological and decision-making processes better than any classical model \cite{Pot09, Bus12, Pot22}. A large and growing body of research has provided significant evidence speaking in favour of this hypothesis \cite{Atm10, Con09, Yuk10, Tru11, deC13, Mar13, Aer14, Khr14, Con15, Bro17, Gro17, Ben18, Khr18, Ros21, Aer22, Kov22, Oza20}.

In Figure~\ref{Fig1_1}b that was rendered using the physical ray-tracing software POV-Ray, we illustrate how the principle of projective qubit measurement can be generalised to the Necker cube. The two-dimensional (bottom) images in Figure~\ref{Fig1_1}b are the shadows cast by the three-dimensional cubes. However, while the three-dimensional cubes are visually different, the shadows cast by them are identical. Yet, the shadows are an ambiguous Necker cube with the alternating left and right faces (this can be seen by observing them for a 5--10\,seconds; some observers may also need to blink to notice the optical illusion \cite{Ang20}). Drawing an analogy with the projective measurement pictured in Figure~\ref{Fig1_1}a, in Figure~\ref{Fig1_1}b we consider the shadows as a qubit-like superposition of the two fundamental perceptual states of the cube and we virtually project these images back to the three-dimensional space to obtain an unambiguous (either $|0\rangle$ or $|1\rangle$ basis state) image of the cube.
    
It is noteworthy that the application of the concepts of qubit and superposition does not imply the existence of quantum processes in a biological brain. In fact, the analogy with a qubit serves as a mathematical model that can adequately describe the experimental data. At the same time, an ultimate verification of the accuracy of the quantum models is not practicable due to technical immaturity and high-cost of quantum computers and adjacent technologies. Yet, from the neurobilogical point of view, an idealised experiment would also include measurements conducted with a brain-computer interface that can decipher the human `thoughts'. Clearly, such complex tests are not yet feasible and they also raise ethics concerns.

Subsequently, much of the current research in this area has focused on artificial neural network modelling and digital twins of perception of optical illusions \cite{Hop82, Rum87, Ino94, Gae98, Noe12, Ara20, Bat22}. Some of these works have employed experimental electroencephalogram (EEG) and magnetoencephalography (MEG) data as the signals that are processed using a neural network model and then classified and correlated with experimental perceptual states of ambiguous figures \cite{Gae98, Noe12, Ara20, Bat22}. In turn, the works \cite{Rum87, Kan89, Ino94} have focused on the analysis of the dynamics of perception of the images of the ambiguous figures using neural network architectures that exhibit a chaotic behaviour. However, the results obtained in Ref.~\cite{Ino94} reproduce the results obtained in classical models of bistable perception, i.e. they do not predict any superposition of the two possible perceptual states of the Necker cube (quantum-mechanical models of cognition and perception had not been widely accepted when the paper \cite{Ino94} was published). On the other hand, although the paper \cite{Kan89} does not discuss the perception of ambiguous figures, the neural network model proposed in it reveals a possibility of a superposition of two states in principle.

In this present work, we construct a deep neural network that uses a quantum random generator to define the weights of the neural connections and we exploit it to model the perception of the ambiguous figures. We demonstrate that the so-designed computational algorithm reproduces the hypothesised superposition of the possible perceptual states of the Necker cube. We also show that these results agree with the predictions of a recently proposed quantum oscillator model of optical illusions \cite{Mak23_quantum}.  

\section{Deep Neural Network Algorithm}
\begin{figure}[t]
 \includegraphics[width=0.99\textwidth]{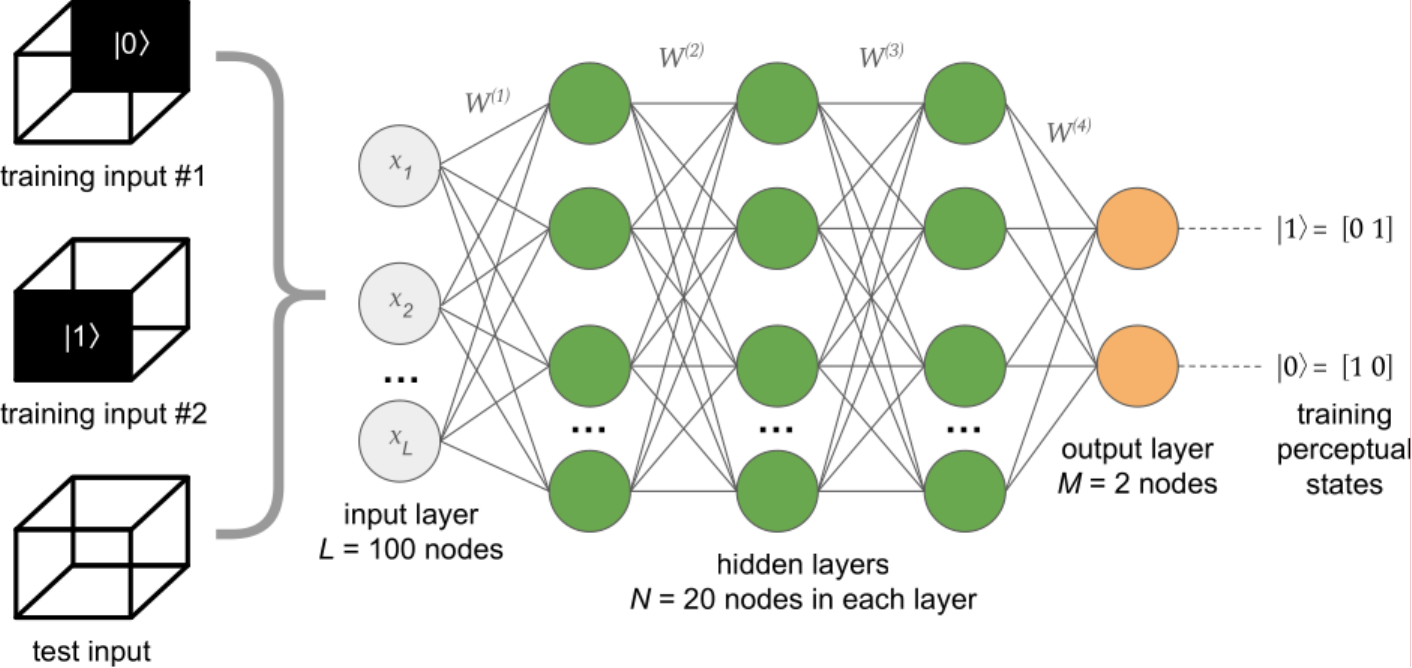}
 \caption{Sketch of the deep neural network architecture used to model the perception of the Necker cube. The network consists of an input layer, three hidden layers and an output layer that has two nodes. The labels $W^{(n)}$ with $n=1\dots4$ denote the matrices of the weights of the network connections. The network is trained using the images of the Necker cube with the shaded front and rear faces that correspond to the $|0\rangle=[0~1]$ and $|1\rangle=[1~0]$ training states, respectively. The characters `$|0\rangle$' and `$|1\rangle$' are not a part of the training images. The test input is an image of the ambiguous Necker cube. All input images consist of a total of 100 pixels. The individual pixels of each image form an input vector $x_j$ with $j=1\dots100$.\label{Fig2}}
\end{figure}

The architecture of the neural network used in this work is illustrated in Figure~\ref{Fig2}. The network consists of an input layer that has $L=100$ input nodes, three hidden layers each of which has $N=20$ nodes and an output layer that has $M=2$ output nodes that are used to classify the perceptual state of the Necker cube. The weights of the connections of the network are updated using a cross entropy-driven back-propagation algorithm \cite{Goo16, Kim17}. The learning rate parameter used in all computations is $\alpha=0.01$.

The activation function of the nodes of the hidden layers are represented by the Rectified Linear Unit (ReLU) that can be defined as \cite{Goo16, Kim17} 
\begin{equation}
  \phi_{ReLU}(x_j) = 
  \begin{cases}
      x, & \text{$x_j>0$}\\
      0, & \text{$x_j\leq0$}
    \end{cases}\,,
  \label{eq:ReLU}
\end{equation}
where $j=1 \dots L$ is the index denoting the sequential number of the input node and $x_j$ is the output from this node. As the activation function of the output nodes we choose the Softmax function that accounts not only for the weighted sum of the inputs to the given node but also for the inputs to the other output nodes \cite{Goo16, Kim17}. This function is 
\begin{equation}
  \phi_{smax}(v_i) = \frac{\exp(v_i)}{\sum_{k=1}^{M} \exp(v_k)}\,,
  \label{eq:softmax}
\end{equation}
where $v_i$ is the weighted sum of the input signals to the $i$th output node and $M$ is the total number of the output nodes. The use of Eq.~(\ref{eq:softmax}) enables satisfying the probability normalisation condition $\sum_{k=1}^{M}\phi_{smax}(v_k)=1$.

The network is trained using the following procedure \cite{Kim17}:
\begin{enumerate}[leftmargin=1in]
\item Construct two output nodes that correspond to $|0 \rangle = \left[1~0\right]$ and $|1 \rangle = \left[0~1\right]$ perceptual states of the Necker cube;
\item Initialise the weights of the neural network in the range from --1 to 1 using a random number generator;
\item Enter the input data $x_j$ and the corresponding training data $d_i$ that encode the perceptual states of the Necker cube (the top and the middle illustrations on the left of Figure~\ref{Fig2});
\item Calculate the error $e_i$ between the output $y_i$ and target $d_i$ as $e_i=d_i-y_i$;
\item Propagate the output $\delta_i=e_i$ in the backward direction of the network and compute the respective parameters $\delta_i^{(n)}$ of the hidden nodes using the equations $e_i^{(n)} = W^{{(n)}^\top}\delta_i$ and $\delta_i^{(n)} =\phi_{ReLU}^\prime\left(v_i^{(n)}\right)e_i^{(n)}$, where the index $n$ denotes the sequential number of the hidden layer, prime denotes the derivative of the activation function and $W^\top$ is the transpose of the matrix of weights corresponding to each relevant layer of the network.
\item Repeat Step~5 until the back-propagation algorithm reaches the first hidden layer;
\item Update the weights using the learning rule $w_{ij}^{(n)}:=w_{ij}^{(n)}+\Delta w_{ij}^{(n)}$, where $w_{ij}^{(n)}$ are the weights between an output node $i$ and input node $j$ of the $n$th layer and $\Delta w_{ij}^{(n)} =\alpha \delta_i^{(n)} x_j$;  
\item Repeat Steps 4--7 for all values of the training data set.
\item Repeat Steps 4--8 until the neural network is trained with a desired accuracy.
\end{enumerate}

The exploitation process essentially reproduces Steps 1--3 \cite{Kim17}. We established that it suffices to use 1000~epochs to obtain convergent results in all calculations.

The physical processes underpinning the dynamics of switching between the perceptual states of ambiguous figures remains a subject of debate \cite{Leh95, Kor05}. One of the currently accepted theories suggests that the switching is likely to be explained by chaotic processes observed in nonlinear dynamical systems \cite{Kan89, Ino94, Sak95, Shi10, Che23}. Indeed, broadly speaking, the brain is a dynamical system that that exhibits a complex nonlinear and chaotic behaviour at multiple levels \cite{Bab88, McK94, Kor03}. Subsequently, it is plausible that certain highly nonlinear and chaotic physical system can approximate the behaviour of a brain at least in principle \cite{Mak23_review}. 
\begin{figure}[t]
 \includegraphics[width=0.99\textwidth]{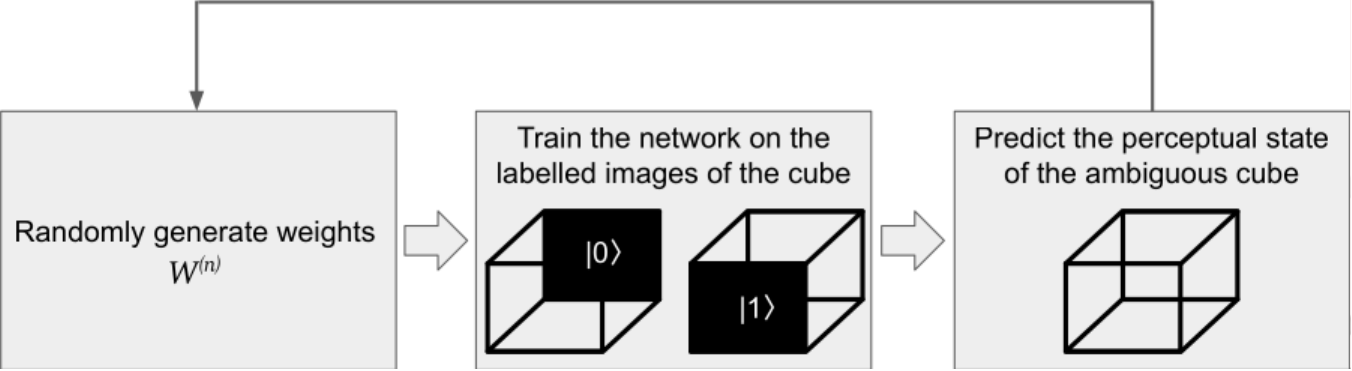}
 \caption{Sketch of the recurring computation procedure that involves the generation of the connection weights using a random generator, training of the network and its exploitation to predict the perceptual state of the Necker cube. The characters `$|0\rangle$' and `$|1\rangle$' are not a part of the training images.\label{Fig3}}
\end{figure}

To implement a chaotic dynamical behaviour in our model, we employ a quantum-physical generator of random numbers \cite{Sym11, Haw15} to define the matrices $W^{(n)}$ that contain the weights of the connections of the neural network. Unlike the output of a pseudo-random generator such as the one described in Ref.~\cite{Rei92}, a quantum generator produces truly random numbers \cite{Sym11, Haw15}. In our model, this property implies that the neural network is not biased towards one of the possible perceptual states of the Necker cube and that its predictions do not repeat in time \cite{Her89, Fan18}. Furthermore, as with the purely classical neural network models \cite{Kan89, Ino94}, our quantum random generator based neural network exhibits a truly chaotic dynamical behaviour \cite{Bru03} and, therefore, can be considered to be a chaos-driven system \cite{Kan89, Ino94}. 

As illustrated in Figure~\ref{Fig3}, we first randomly generate $W^{(n)}$, then we train the network on the data corresponding to the Necker cubes with the shaded faces and then we exploit the trained neural network to predict the perceptual state of the ambiguous Necker cube. This procedure is repeated in a loop to plot the perceived states of the cube as a function of time. 

\subsection{Results:~Predictions of the Neural Network Model}
\begin{figure}[t]
 \includegraphics[width=0.99\textwidth]{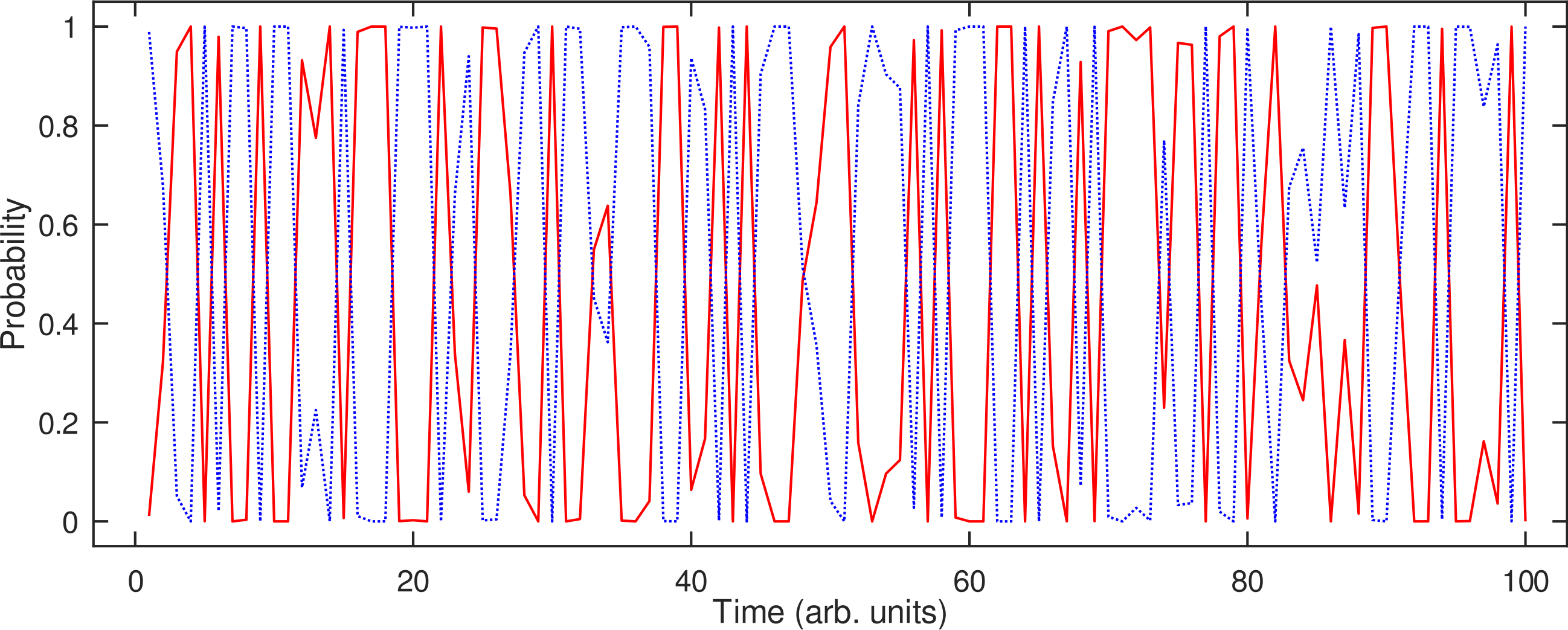}
 \caption{Perceptual switching curves simulated by the neural network model using quantum-random neural connection weights. The data produced by the two output nodes of the network are plotted using the solid and dotted curves, respectively. The data points with the probability $P_{|0\rangle}=0$ or $P_{|1\rangle}=1$ correspond to the fundamental perceptual states of the Necker cube. The remaining data points are in a superposition of the states $|0\rangle$ and $|1\rangle$ with $P_{|0\rangle}+P_{|1\rangle}=1$.\label{Fig5}}
\end{figure}

Figure~\ref{Fig5} shows the prediction by the neural network model obtained as a result of 100~consecutive runs of the algorithm outlined in Figure~\ref{Fig3}. The states of the output nodes of the neural network were recorded at end of each computational run and the respective results were plotted as a function of time (in arbitrary units). Therefore, every pair of data points that constitute the curves in Figure~\ref{Fig5} was obtained using a unique initial set of neural weights $W^{(n)}$ obtained from a truly random quantum-physical system \cite{Sym11, Haw15}.

We can observe a time-dependent switching between the two possible classical perception states of the cube that correspond to the probability values 0 and 1 on the $y$-axis of Figure~\ref{Fig5}. Importantly, the pattern of the switching between one perceptual state to another is not abrupt, as often depicted in the literature and schematically shown in Figure~\ref{Fig1}b of this paper, but gradual. Thus, the data produced by the neural network model speak in favour of plausibility of the previous theoretical results \cite{Bus12} and experimental evidence \cite{Gae98, Pia17, Joo20} demonstrating that the actual perception state is a superposition of the two fundamental states $|0\rangle$ and $|1\rangle$ of the Necker cube.

Similar results were obtained using a quantum oscillator model of perception of ambiguous figures. In the following section, we overview the algorithm of that model and then compare its predictions with the result shown in Figure~\ref{Fig5}.

\section{Quantum Oscillator Model of Perception of Ambiguous Figures}
\begin{figure}
 \includegraphics[width=0.99\textwidth]{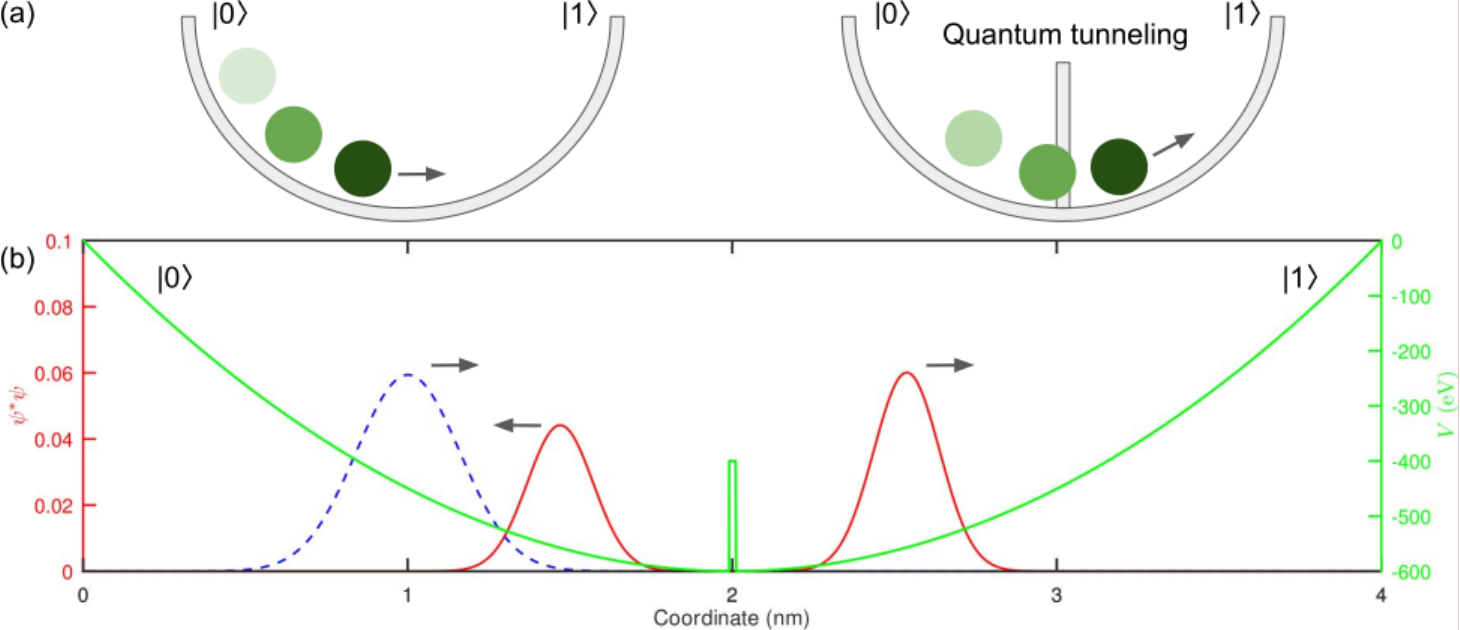}
 \caption{{\bf(a)}~An electron trapped in a parabolic well behaves as a harmonic oscillator and it can pass through a barrier due to the quantum tunnelling effect. {\bf(b)}~Illustrative example (numerical simulation) of quantum tunnelling through a barrier. The dashed line denotes an snapshot of the incident Gaussian pulse. The solid line denotes the snapshots of the portions of the pulse that are reflected from and transmitted through the barrier (the green line, right $y$-axis). The labels $|0\rangle$ and $|1\rangle$ correspond to the perceptual states of the Necker cube.\label{Fig4}}
\end{figure}

We model the dynamics of perception of the Necker cube using a harmonic motion of an electron trapped in a parabolic potential well (Fig.~\ref{Fig4}a). This model is inspired by the quantum-mechanical approach to human cognition proposed in Ref.~\cite{Bus12} and it captures the complex pattern of perception of the Necker cube \cite{Mak23_quantum}.

A classical mechanics counterpart of this model is a small ball that rolls back and forth inside a bowl. While the ball does not have enough energy to surmount or penetrate a physical barrier inside the bowl, the electron may pass through the barrier due to the quantum tunnelling effect (Fig.~\ref{Fig4}a).

We model the quantum tunnelling effect by solving the Schr{\"o}dinger equation in a one-dimensional space \cite{Gri04}
\begin{equation}
  \label{eq:SE}
  i\hbar\frac{\partial \psi(x,t)}{\partial t}=\left[-\frac{\hbar^2}{2m}\frac{\partial^2}{\partial x^2} + V(x)\right]\psi(x, t)\,, 
\end{equation}
where $\psi(x, t)$ is a wave function, $i$ is the imaginary unit, $m$ is the mass of the electron, $\hbar$ is Plank's constant and $V(x)$ is the parabolic potential well profile. We numerically solve Eq.~(\ref{eq:SE}) using a finite-difference time-domain (FDTD) method \cite{Sullivan} that represents the wave function as $\psi(x, t)=\psi_{re}(x, t)+i\psi_{im}(x, t)$. We obtain
\begin{gather}
  \label{eq:Eq3}
  \frac{\partial \psi_{re}(x, t)}{\partial t} = -\frac{\hbar}{2m}\nabla^2 \psi_{im}(x, t)+\frac{1}{\hbar}V(x)\psi_{im}(x, t) \\ \nonumber
\frac{\partial \psi_{im}(x, t)}{\partial t} = \frac{\hbar}{2m}\nabla^2 \psi_{re}(x, t)-\frac{1}{\hbar}V(x)\psi_{re}(x, t)\,. 
\end{gather}

Representing the coordinate $x$ and time $t$ as the vectors of discrete elements $x_k = k \Delta x$ and $t_n = n \Delta t$, respectively, where $k$ and $n$ are integer numbers, and applying the Courant stability criterion \cite{Sullivan}, we define
\begin{equation}
  \label{eq:Eq4}
  \Delta t = \frac{1}{8}\frac{2m}{\hbar} (\Delta x)^2\,. 
\end{equation}
Thus, a spatio-temporally discretised Eq.~(\ref{eq:Eq3}) becomes 
\begin{gather}
  \label{eq:Eq5}
  \psi_{re}^{n}(k) = \psi_{re}^{n-1}(k)-\frac{1}{8}\left[\psi_{im}^{n-1/2}(k+1)-2\psi_{im}^{n-1/2}(k)+\psi_{im}^{n-1/2}(k-1)\right]+\frac{\Delta t}{\hbar}V(k)\psi_{im}^{n-1/2}(k)\\ \nonumber
  \psi_{im}^{n}(k) = \psi_{im}^{n-1}(k)+\frac{1}{8}\left[\psi_{re}^{n-1/2}(k+1)-2\psi_{re}^{n-1/2}(k)+\psi_{re}^{n-1/2}(k-1)\right]-\frac{\Delta t}{\hbar}V(k)\psi_{re}^{n-1/2}(k)\,. 
\end{gather}

We model the electron as a Gaussian energy wave packet:
\begin{gather}
  \label{eq:Eq6}
  \psi_{re}^{0}(k) = \exp\left(-0.5\left(\frac{k-k_0}{\sigma}\right)^2\right) \cos\left(\frac{2\pi(k-k_0)}{\lambda}\right)\\ \nonumber
  \psi_{im}^{0}(k) = \exp\left(-0.5\left(\frac{k-k_0}{\sigma}\right)^2\right) \sin\left(\frac{2\pi(k-k_0)}{\lambda}\right)\,, 
\end{gather}
where $\lambda$ is the wavelength, $\sigma$ is the width of the Gaussian pulse and $k_0$ is the spatial coordinate of origin of the pulse. The amplitudes of the wave functions are normalised as
\begin{equation}
  \label{eq:Eq7}
  \int_{-\infty}^{\infty} \psi^{*}(x)\psi(x) \,dx = 1\,. 
\end{equation}
The probabilities of funding the electron is the $|0\rangle$ and $|1\rangle$ regions of the potential well are calculated as 
\begin{gather}
  \label{eq:Eq8}
  P_{|0\rangle}=\int_{-\infty}^{x_{centre}} \psi^{*}(x)\psi(x) \,dx \\ 
  P_{|1\rangle}=\int_{x_{centre}}^{\infty} \psi^{*}(x)\psi(x) \,dx\,, 
\end{gather}
where $P_{|0\rangle}+P_{|1\rangle}=1$.
\begin{figure}
 \includegraphics[width=0.99\textwidth]{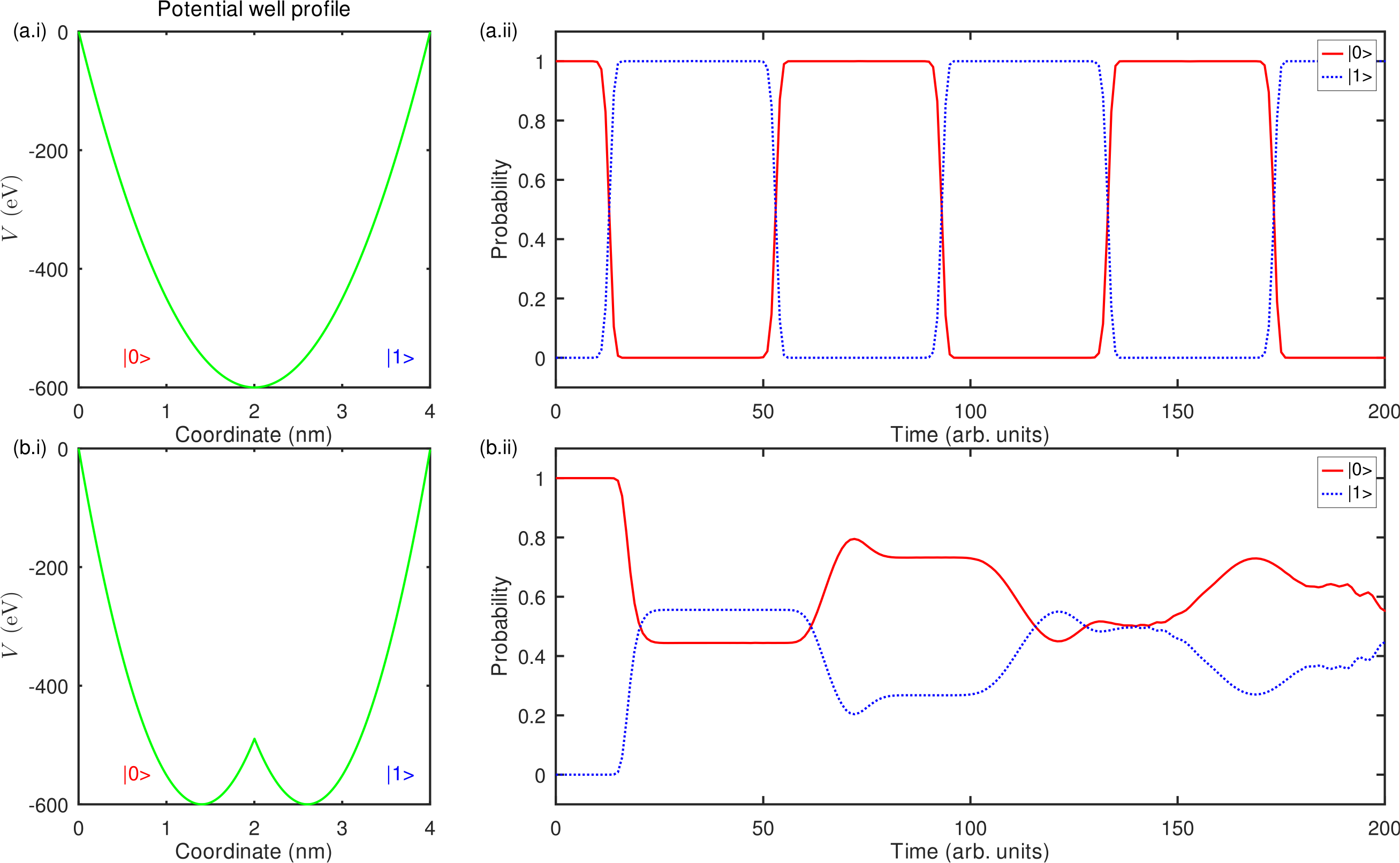}
 \caption{{\bf(a)}~A single parabolic potential well model and the Necker cube perception switching predicted by it. {\bf(b)}~The respective double-parabolic model with a barrier and its predictions. In both panels, the labels $|0\rangle$ and $|1\rangle$ denote the fundamental perceptual states of the Necker cube. The time units used in this figure are different from those in Figure~\ref{Fig5}.\label{Fig6}}
\end{figure}

Using the model parameters $\Delta x=0.1\times10^{-11}$\,m, $\lambda=1.6\times10^{-10}$\,m and $\sigma=1.6\times10^{-10}$\,m, in Figure~\ref{Fig4}b we present the results of modelling of the electron tunnelling through a potential barrier. Calculating the modulus square of the wave function, we obtain the probability density of finding the electron at a certain position in the parabolic potential well. We can see that one part of the incident wave packet is reflected from the barrier but another part is transmitted through it. We label the left and right side of the parabolic potential well as $|0\rangle$ and $|1\rangle$ and associate them with the possible perceptual states of the Necker cube. In this particular demonstration simulation scenario, we obtain $P_{|0\rangle}=0.35$ and $P_{|1\rangle}=0.65$.

\subsection{Results:~Predictions of the Quantum Oscillator Model}
Figure~\ref{Fig6} shows the results produced by the quantum oscillator model. We consider a single parabolic well (Figure~\ref{Fig6}a.i) and a double-parabolic well with a barrier formed by two overlapping parabolic wells (Figure~\ref{Fig6}b.i). Assuming that the energy packet that represents the electron originates from the left side of the potential well (this corresponds to a visual cue to the cube orientation \cite{Bus12}), we simulate the dynamics of the oscillator in the time interval from 0 to 200 arbitrary units (these arbitrary units are different from those used in the neural network model). The result of this simulation is plotted in Figure~\ref{Fig6}a.ii, where the probability of finding the electron in the $|0\rangle$ and $|1\rangle$ regions of the potential well are denoted by the solid and dotted curves, respectively. The result of the simulation of the double-parabolic well is presented in Figure~\ref{Fig6}b.ii.

\section{Discussion}
\subsection{Neural Network Model versus Quantum Oscillator Model}
In Figure~\ref{Fig6}a.ii, we can see that the quantum oscillator model with a single potential well predicts a periodic switching between the two fundamental perceptual states, with a quick but not instantaneous change from one fundamental perceptual state to another. This result is similar to the prediction of the previous quantum-mechanical models proposed in Refs.~\cite{Atm10, Bus12} and it implies the existence of a superposition of the fundamental perceptual states. A qualitatively similar results is predicted by the neural network, which can be seen in Figure~\ref{Fig5} in the time intervals from approximately $T=5$ to $T=20$.

Furthermore, a periodic switching between the two fundamental perceptual states predicted by the neural network model alternates with the periods of irregular switching between these states (e.g.~from $T=80$ to approximately $T=90$ in Figure~\ref{Fig5}). This behaviour is qualitatively reproduced by the quantum oscillator model that uses the double-periodic well with the barrier (Figure~\ref{Fig6}b.ii). We note that the time units used in the quantum oscillator model are different from those used in the neural network model, which means that the timescale of alternations between the perceptual states is different in these two models. This difference is inconsequential for the current discussion and, if needed, it can be eliminated using a different profile of the parabolic potential wells.     

Thus, we conclude that the quantum oscillator model can reproduce the predictions of the neural network model provided that the outputs of the single and double parabolic potential well oscillators are combined together, which can be done, for example, by coupling them into a chain oscillator. While the discussion of an implementation of this approach is beyond the scope of this paper, the similarity of the outputs of the neural network model and the quantum oscillator model has a clear physical meaning: both models are dynamical systems that operate according to the fundamental laws of quantum mechanics \cite{Koc21, Dou22, Koc22}.  

Moreover, the results obtained using the neural network model speak in favour of the hypothesis that originates from the quantum oscillator model and that suggests that the phase change in the response of the dynamical system has effect of eye blinking, an action known to induce a reversal of the perceptual state of the Necker cube \cite{Kor12, Ang20, Mak23_quantum}. Although this hypothesis has not been verified yet, it is known that the dynamics of eye blinks can be studied using the methods developed to investigate highly nonlinear and chaotic processes \cite{Ham12, Pap17, Har18}. Hence, since the neural network model employs data produced by a generator of truly random numbers, its predictions should be consistent with the dynamics of the eye blink \cite{Her89, Ric07}.   

\subsection{Potential Applications in Artificial Intelligence and Virtual Reality Systems}
The proposed neural network algorithm can be used as a model of optical illusions in film-making, architecture design and game development \cite{Fis67, Smi84, game2}. For example, the video game Superliminal uses forced perspective techniques that manipulate human visual perception to make an object appear larger or smaller than it actually is \cite{game1}. The same algorithm can be used in an advanced machine vision system intended to simulate the human perception and decision-making. In particular, the so-designed machine vision system may be tasked to play a video game such as Superliminal and its actions can be compared with the actions of a human operator, providing a valuable feedback for engineers, neuroscientists and psychologists.  

The neural network model of optical illusions can be used to study the impact of weightlessness on the ability of astronauts to undertake complex tasks during and after spaceflights. On Earth, the majority of observers of ambiguous figures such as the Necker cube perceive one interpretation more often than the other. However, in weightlessness, this asymmetry gradually disappeared and, after spending several months in orbit, both interpretations have the same occurrence \cite{Yam06, Cle17}.

The operation of unmanned aerial vehicle (UAVs), commonly known as drones, is another area where models of optical illusions may help extend the abilities of both humans and AI. For example, at present the skills of human race drones operators significantly exceed the performance of the most advanced machine vision algorithms \cite{Pfe21}. A better understanding of the ability of human pilot to select appropriate motor commands from highly dynamic visual information may provide key insights for solving current challenges in vision-based autonomous navigation. 

Yet, the neural network model of optical illusions can be used to validate certain neuroscience and psychological perception theories that are complimentary to the quantum mind hypothesis \cite{Ber00, Doy07}. According to some mainstream theories, our subjective perception of the world is unitary coherent \cite{Fox12}. Here, unitary means that we perceive one interpretation at a time (e.g.~one of the two possible states of the Necker cube) rather than a blur of the possible interpretations (i.e.~we never see the two possible states of the cube together). In turn, coherent means that we perceive scenes that do not contain contradictory parts (e.g.,~we do not see a part of one cube and a part of another one at the same time).

However, such an intuitive approach contradicts the theories of optimal decision-making and Bayesian brain \cite{Ber00, Doy07}. These theories suggest that an optimal decision can be made only integrating the utility of all actions while considering all possible interpretations of sensory data.

To verify these alternative theories, a video game involving two scenarios was designed \cite{Fox12}, where the players were first trained in a visually unambiguous scenario and then they played the same game but in an optical illusion scene that involved an image of the Necker cube. The proposed neural network model can be integrated with that game to address the weaknesses of the experiment identified in Ref.~\cite{Fox12}.    

\section{Conclusions}
This paper demonstrates the potential of a deep neural network algorithm powered by a quantum random number generator to simulate the human perception of optical illusions exemplified by the Necker cube. The results produced by the model indicate that observers are likely to perceive a superposition of the fundamental perceptual states of the cube.

This finding aligns with the emerging psychology theories suggesting that certain psychological phenomena can be adequately described using such quantum-mechanical concepts as qubit, superposition of states and projective measurement. In particular, we compared the results produced by the neural network with the predictions of a recently proposed quantum oscillator model of optical illusions and we established that both models consistently predict a qubit-like superposition of perceptual states.      

The proposed neural network model can be used in various AI systems ranging from video games and virtual reality and metaverse products, also being a useful tool for psychological and neuroscience studies. It can also be utilised to train astronauts and operators of UAVs to perform in visually challenging environments. 

\vspace{6pt}

\authorcontributions{All results presented in this article were obtained by the author. The author conceived the idea of writing this article and prepared the manuscript.}

\funding{Not applicable.}

\institutionalreview{Not applicable.}

\informedconsent{Not applicable.}

\dataavailability{The computational code of the neural network used in this work can be found at \url{https://github.com/IvanMaksymov/DeepNeuralNecker}.}

\acknowledgments{The author acknowledges useful discussions with Professor Ganna Pogrebna.}

\conflictsofinterest{The author declares no conflict of interest.}

\abbreviations{Abbreviations}{
The following abbreviations are used in this manuscript:\\

\noindent 
\begin{tabular}{@{}ll}
 artificial Intelligence &AI\\
 electroencephalogram &EEG\\
 finite-difference time-domain &FDTD\\
 magnetoencephalography &MEG\\
 Rectified Linear Unit &ReLU\\
 unmanned aerial vehicle &UAV\\
  
\end{tabular}
}


\begin{adjustwidth}{-\extralength}{0cm}
\reftitle{References}


\externalbibliography{yes}
\bibliography{refs}

\end{adjustwidth}
\end{document}